\documentclass[twocolumn,pre,showpacs]{revtex4}
\usepackage{graphicx}
\usepackage{amsmath}
\usepackage{amsfonts}
\usepackage{amssymb}
\usepackage{color}
\usepackage{float}
\usepackage{mathrsfs}

\begin{document}
\title{Rogue waves with rational profiles in unstable condensate and its solitonic model}

\author{D.\,S.~Agafontsev$^{1,2}$}
\email{dmitrij@itp.ac.ru}
\author{A.\,A.~Gelash$^{2,3}$}

\affiliation{\textit{$^1$ P.P. Shirshov Institute of Oceanology of RAS, 117997 Moscow, Russia.\\
$^2$ Skolkovo Institute of Science and Technology, 121205 Moscow, Russia.\\
$^3$ Institute of Automation and Electrometry of SB RAS, 630090 Novosibirsk, Russia.}}

\begin{abstract}
In this brief report we study numerically the spontaneous emergence of rogue waves in (i) modulationally unstable plane wave at its long-time statistically stationary state and (ii) bound-state multi-soliton solutions representing the solitonic model of this state [Gelash et al, PRL 123, 234102 (2019)]. 
Focusing our analysis on the cohort of the largest rogue waves, we find their practically identical dynamical and statistical properties for both systems, that strongly suggests that the main mechanism of rogue wave formation for the modulational instability case is multi-soliton interaction. 
Additionally, we demonstrate that most of the largest rogue waves are very well approximated -- simultaneously in space and in time -- by the amplitude-scaled rational breather solution of the second order. 
\end{abstract}

\maketitle


\section{Introduction}
\label{Sec:Intro}

The phenomenon of rogue waves (RWs) -- unusually large waves that appear suddenly from moderate wave background -- was intensively studied in the recent years. 
A number of mechanisms were suggested to explain their emergence, see e.g. the reviews~\cite{kharif2003physical,dysthe2008oceanic,onorato2013rogue}, with the most general idea stating that RWs could be related to breather-type solutions of the underlying nonlinear evolution equations~\cite{dysthe1999note,osborne2010nonlinear,shrira2010makes}. 
Currently, ones of the most popular models for RWs are the Peregrine rational breather~\cite{peregrine1983water} and the higher-order rational breather~\cite{akhmediev2009rogue} solutions of the one-dimensional nonlinear Schr{\"o}dinger equation (1D-NLSE) of the focusing type,
\begin{equation}\label{NLSE}
	i\psi_t + \psi_{xx} + |\psi|^2 \psi = 0.
\end{equation}
These rational breathers represent a family of localized in space and time algebraic solutions, which evolve on a finite background and lead to three-fold, five-fold, seven-fold, and so on, increase in amplitude at the time of their maximum elevation. 
Taking specific and carefully designed initial conditions, they were reproduced in well-controlled experiments performed in different physical systems~\cite{kibler2010peregrine,chabchoub2011rogue,bailung2011observation,chabchoub2012super,chabchoub2012observation}.

The 1D-NLSE is integrable in terms of the \textit{inverse scattering transform} (IST), as it allows transformation to the so-called \textit{scattering data}, which is in one-to-one correspondence with the wavefield and, similarly to the Fourier harmonics in the linear wave theory, changes trivially during the motion. 
Thanks to its properties, the scattering data can be used to characterize the wavefield. 
For spatially localized case, the scattering data consists of the discrete (solitons) and the continuous (nonlinear dispersive waves) parts of eigenvalue spectrum, calculated for specific auxiliary linear system. 
For strongly nonlinear wavefields, such as the ones where emergence of rational breathers can be expected, the solitons provide the main contribution to the energy~\cite{novikov1984theory} and should therefore play the dominant role in the dynamics. 
In particular, as has been recently demonstrated in~\cite{gelash2019bound}, the modulationally unstable plane wave (the condensate) at its long-time statistically stationary state can be accurately modeled (in the statistical sense) with a certain soliton gas, designed to follow the solitonic structure of the condensate. 
The latter naturally raises a question of whether there is a difference between the RWs emerging in the two systems. 
Indeed, in a soliton gas all RWs are multi-soliton interactions by construction. 
Hence, if there is no significant difference, then we can draw a hypothesis that for the asymptotic stationary state of the MI (and, possibly, for other strongly nonlinear wavefields) the main mechanism of RW formation is interaction of solitons. 

With the present paper, we contribute to the answer on this question by summarizing our observations of RWs for both systems. 
Specifically, we compute time evolution for $1000$ random realizations of the noise-induced MI of the condensate and also for $1000$ random realizations of $128$-soliton solutions modeling the asymptotic state of the MI. 
For each realization, we analyze one largest RW emerging in the course of the evolution, thus focusing our analysis on the largest RWs. 
For both systems, we observe practically identical dynamical and statistical properties of the collected RWs. 
In particular, most of the RWs turn out to be very well approximated -- simultaneously in space and in time -- by the amplitude-scaled rational breather solution (RBS) of the second order. 
By measuring the deviation between the RWs and their fits with RBS as an integral of the difference in the $(x,t)$-space, we find that, in general, the larger the maximum amplitude of the RW, the better its convergence to the RBS of the second order (RBS2). 
The collected RWs for the two systems turn out to be identically distributed by their maximum amplitude and deviation from the RBS2. 
Additionally, we demonstrate that the observed quasi-rational profiles appear already for synchronized three-soliton interactions and discuss the next steps in the ongoing research of the RW origin. 

Note that in the present paper we consider solutions of the 1D-NLSE for three different types of boundary conditions: the MI of the condensate for which we use the periodic boundary, the multi-soliton solutions with vanishing border conditions and the RBS having constant border conditions at infinity. 
Globally, these solutions are fundamentally different, and the different border conditions require application of separate IST techniques, see e.g.~\cite{novikov1984theory,belokolos1994algebro,osborne2010nonlinear,bobenko2011computational}. 
For instance, formally our MI case corresponds to finite-band scattering data. 
However, the characteristic widths of the structures (RWs, solitons, RBS) are small compared to the sizes of the studied wavefields, so that the eigenvalue bands are very narrow and we neglect their difference from solitons. 
The similar idea was suggested in~\cite{el2001soliton}, where, vice versa, the soliton gas was considered as a limit of finite-band solutions. 
Effectively, we assume that formation of a RW, as a local phenomenon, represents a similar process for all three cases of border conditions. 
As we demonstrate in the paper, this assumption is supported by the presented results, that raises an important problem that we leave for future studies -- explanation of how the three models may exhibit locally similar nonlinear patterns.

The paper is organized as follows. 
In the next Section we describe our numerical methods and initial conditions, and also discuss how we approximate a RW with a RBS. 
In Section~3 we summarize our observations. 
Section~4 is devoted to discussion, and the final Section~5 contains conclusions. 


\section{Numerical methods}
\label{Sec:NumMethods}


We solve Eq.~(\ref{NLSE}) in a large box $x\in[-L/2, L/2]$, $L\gg 1$, with periodic boundary conditions using the pseudo-spectral Runge-Kutta fourth-order method in adaptive grid with the grid size $\Delta x$ set from the analysis of the Fourier spectrum of the solution; see~\cite{agafontsev2015integrable} for detail. 
As an integrable equation, the 1D-NLSE conserves an infinite set of integrals of motion, see e.g.~\cite{novikov1984theory}. 
We have checked that the first ten integrals are conserved by our numerical scheme up to the relative errors from $10^{-10}$ (the first three invariants) to $10^{-6}$ (the tenth invariant) orders. 

Without loss of generality, the initial conditions for the noise-induced MI of the condensate can be written as
\begin{eqnarray}
	\psi|_{t=0} = 1 + \epsilon(x), \label{IC-condensate}
\end{eqnarray}
where $\epsilon(x)$ represents a small initial noise. 
We use statistically homogeneous in space noise with Gaussian Fourier spectrum,
\begin{equation}\label{noise}
	\epsilon(x)=a_{0}\bigg(\frac{\sqrt{8\pi}}{\theta L}\bigg)^{1/2} \sum_{k}e^{-k^{2}/\theta^{2}+i\phi_{k}+ikx},
\end{equation}
where $a_{0}$ is the average noise amplitude in the $x$-space, $k=2\pi m/L$ is the wavenumber, $m\in\mathbb{Z}$ is integer, $\theta$ is the characteristic noise width in the $k$-space and $\phi_{k}$ are random phases for each $k$ and each realization of the initial conditions; the average intensity of such noise equals to $a_{0}^{2}$, $\langle|\epsilon|^{2}\rangle = a_{0}^{2}$. 
For the numerical experiment, we take the box of length $L=256\pi$ and small initial noise, $a_{0}=10^{-5}$, with wide spectrum, $\theta=5$. 
Note that these parameters match those used in~\cite{agafontsev2015integrable}.

To generate the solitonic model of the asymptotic stationary state of the noise-induced MI, we create $128$-soliton solutions with the combination of the dressing method and $100$-digits precision arithmetics as described in~\cite{gelash2018strongly}. 
Each soliton has four parameters: amplitude $a_{j}$, velocity $v_j$, space position $x_{0j}$ and phase $\Theta_{j}$; here $j=1,...,M$, $M=128$, and the one-soliton solution reads as
\begin{eqnarray}
	\psi_{s}(x,t) = a \frac{\exp\bigg[\frac{i v}{2} (x-x_0) + \frac{i}{2} \bigg(a^2-\frac{v^2}{2}\bigg) t + i \Theta\bigg]}{\cosh\frac{a (x-x_0) - a v t}{\sqrt{2}}}. \nonumber
\end{eqnarray}
Following~\cite{gelash2019bound}, we distribute soliton amplitudes according to the Bohr-Sommerfeld quantization rule,
\begin{eqnarray}
	a_{j} = 2\sqrt{1 - \bigg(\frac{j-1/2}{M}\bigg)^{2}}, \label{BS-rule}
\end{eqnarray}
and set soliton velocities to zero, $v_j=0$, using uniformly-distributed soliton phases $\Theta_{j}$ in the interval $[0,2\pi)$ and uniformly-distributed space position parameters $x_{0j}$ in a narrow interval at the center of the computational box. 
Zero velocities mean that these multi-soliton solutions are bound-state. 
For the 1D-NLSE in normalization~(\ref{NLSE}), the Bohr-Sommerfeld rule describes amplitudes for the bound-state solitonic content of the rectangular box wavefield of unit amplitude $\psi=1$ and width $L_{o}=\sqrt{2}\pi M$, calculated with the semi-classical Zakharov-Shabat direct scattering problem, see e.g.~\cite{zakharov1972exact,novikov1984theory,lewis1985semiclassical}. 
The generated $128$-soliton solutions take values of unity order approximately within the interval $x\in [-L_{o}/2, L_{o}/2]$ and remain small outside of it. 
For more detail on the soliton gas, we refer the reader to~\cite{gelash2019bound}, where it has been demonstrated that its spectral (Fourier) and statistical properties match those of the long-time statistically stationary state of the MI.

For the soliton gas, we gather the RWs by simulating the time evolution of the $128$-soliton solutions in the interval $t\in[0,50]$ and then collecting one largest RW for each of the $1000$ realizations of initial conditions. 
For time evolution, we use the same pseudo-spectral Runge-Kutta numerical scheme as for the MI of the condensate, since application of the dressing method with evolving scattering data takes too much computational time and provides the same result. 
The pseudo-spectral scheme uses periodic boundary conditions, so that solution $\psi(x,t)$ needs to be small near the edges of the computational box. 
We achieve this by taking the box of length $L=384\sqrt{2}\pi$, so that our $128$-soliton solutions are of $10^{-16}$ order near its edges and take values of unity order, $|\psi|\sim 1$, only within its central $1/3$ part ($\equiv L_{o}/L$). 

For the MI of the condensate, we collect the RWs similarly, but in the time interval $t\in[174, 200]$. 
From the one hand, the end of this interval is far enough, so that the system is sufficiently close to its asymptotic stationary state, see~\cite{agafontsev2015integrable} where the same initial conditions were used. 
From the other hand, a chance to detect a large RW is higher in larger simulation boxes and if we wait longer. 
To make RW events for the two systems comparable, we impose a restriction $L^{(MI)}\cdot \Delta T^{(MI)} = L^{(SG)}\cdot \Delta T^{(SG)}$ on the lengths $L^{(MI,SG)}$ of the regions where RWs may appear and on the time intervals $\Delta T^{(MI,SG)}$ during which we wait for the largest RW. 
For the soliton gas, the collected RWs appear approximately in the space interval $x\in[-210, 210]$ with practically uniform distribution of their position, so that $L^{(SG)}=420$. 
We believe that this property is connected with the behavior of the ensemble- and time-averaged intensity $I(x)=\langle|\psi(x,t)|^{2}\rangle$, which remains flat $I=1$ inside this interval and starts to deviate from unity at its edges. 
For the MI, the RWs may appear anywhere within the computational box $L^{(MI)} = 256\pi$; together with the observation time for the soliton gas case $\Delta T^{(SG)}=50$, this yields $\Delta T^{(MI)}=26$ and the time interval $t\in[174, 200]$ for the MI.


The rational breather solution of the first order (RBS1) -- the Peregrine breather~\cite{peregrine1983water} -- reads as
\begin{equation}\label{Peregrine1}
	\psi_{p}^{(1)}(x,t)=e^{i t}\bigg[1 - \frac{4(1+2 i t)}{1+2 x^{2}+4 t^{2}}\bigg].
\end{equation}
The RBS of the second order (RBS2) $\psi_{p}^{(2)}$ is too complex and we refer the reader to~\cite{akhmediev2009rogue} where it was first found. 
Both solutions are localized in space and in time, and evolve on a finite background (the condensate). 
For approximation of a RW with a RBS, we use the scaling, translation and gauge symmetries of the 1D-NLSE: indeed, if $u(x,t)$ is a solution of Eq.~(\ref{NLSE}), then $A_{0}e^{i\Theta}\cdot u(\chi,\tau)$, where $\chi = A_{0}(x-x_0)$, $\tau = A_{0}^{2}(t-t_0)$ and $A_{0},\Theta\in\mathbb{R}$, is also a solution. 
Technically, we detect the maximum amplitude $A$ of a RW together with its position $x_0$ and time $t_0$ of occurrence, and also the phase at maximum amplitude $\Theta = \arg\psi(x_0, t_0)$, and then use the scaling coefficient $A_{0} = -A/3$ for the RBS1 and $A_{0} = A/5$ for the RBS2. 

Note that, in general, RBS may have nonzero velocity $v\neq 0$. 
To account its influence, one can make a transformation $u(x,t) \to e^{ivx/2 - iv^2 t/4}\cdot u(x-vt,t)$, which also prompts a simple way to find the velocity. 
Indeed, at the time of the maximum elevation $t_{0}$, a RBS with zero velocity, $v=0$, has constant phase $\arg\psi(x,t_{0})=\mathrm{const}$ in the region between the two zeros closest to the maximum amplitude. 
In contrast, a RBS with nonzero velocity, $v\neq 0$, has constant phase slope, $\arg\psi(x,t_{0}) - ivx/2 = \mathrm{const}$, in the same region. 
Hence, by computing the phase slope one can approximate RWs with RBS of nonzero velocity. 
For all the RW studied in this paper, we have checked that taking into account velocity improves our approximations only very slightly, and for this reason we have decided to use RBS with zero velocity only. 

Also note that in addition to the RBS1 and the RBS2, we have examined approximation with the RBS of the 3rd order~\cite{akhmediev2009rogue}, as well. 
However, we have found that it works worse than either RBS1, or RBS2 for all the $2000$ examined RWs, and thereby excluded it from the analysis.


\section{Rogue waves with rational profiles}
\label{Sec:Results1}

We start this Section with the description of one RW event for the soliton gas case, and then continue with examination of RW properties for both systems -- the noise-induced MI close to its asymptotic stationary state and the soliton gas representing the solitonic model of this state.

\begin{figure*}[t]\centering
	\includegraphics[width=8.8cm]{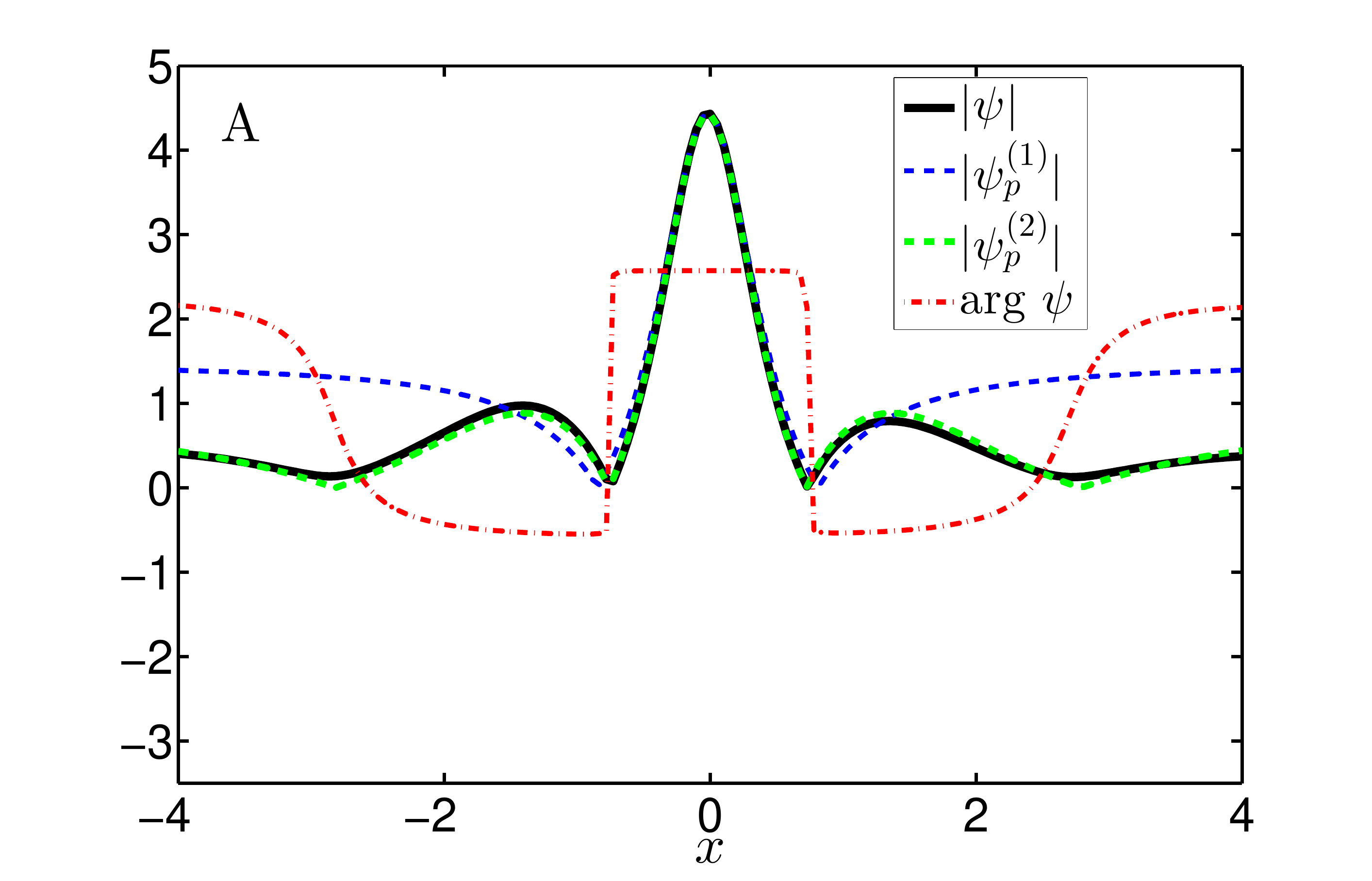}
	\includegraphics[width=8.8cm]{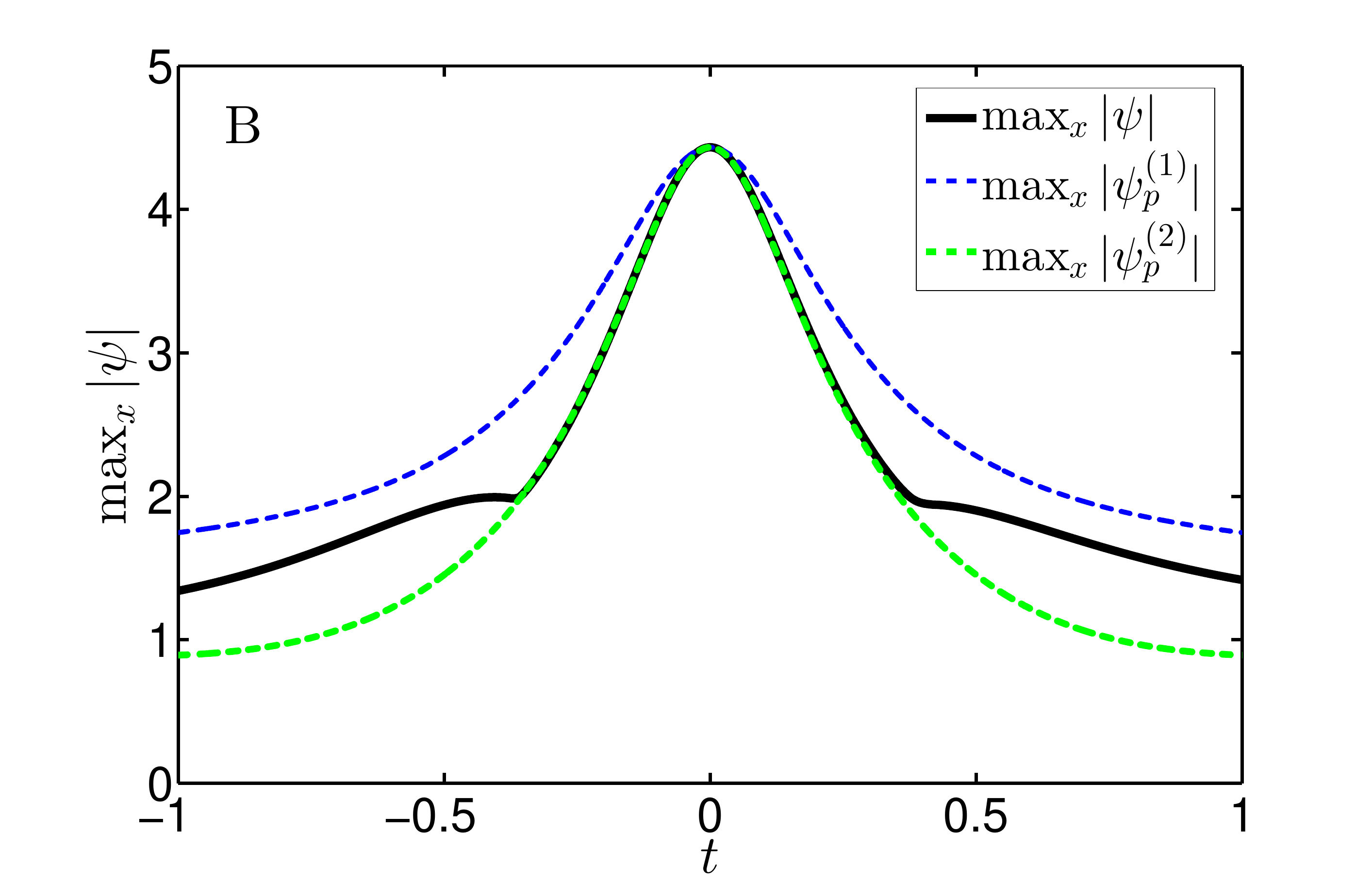}\\
	\includegraphics[width=8.8cm]{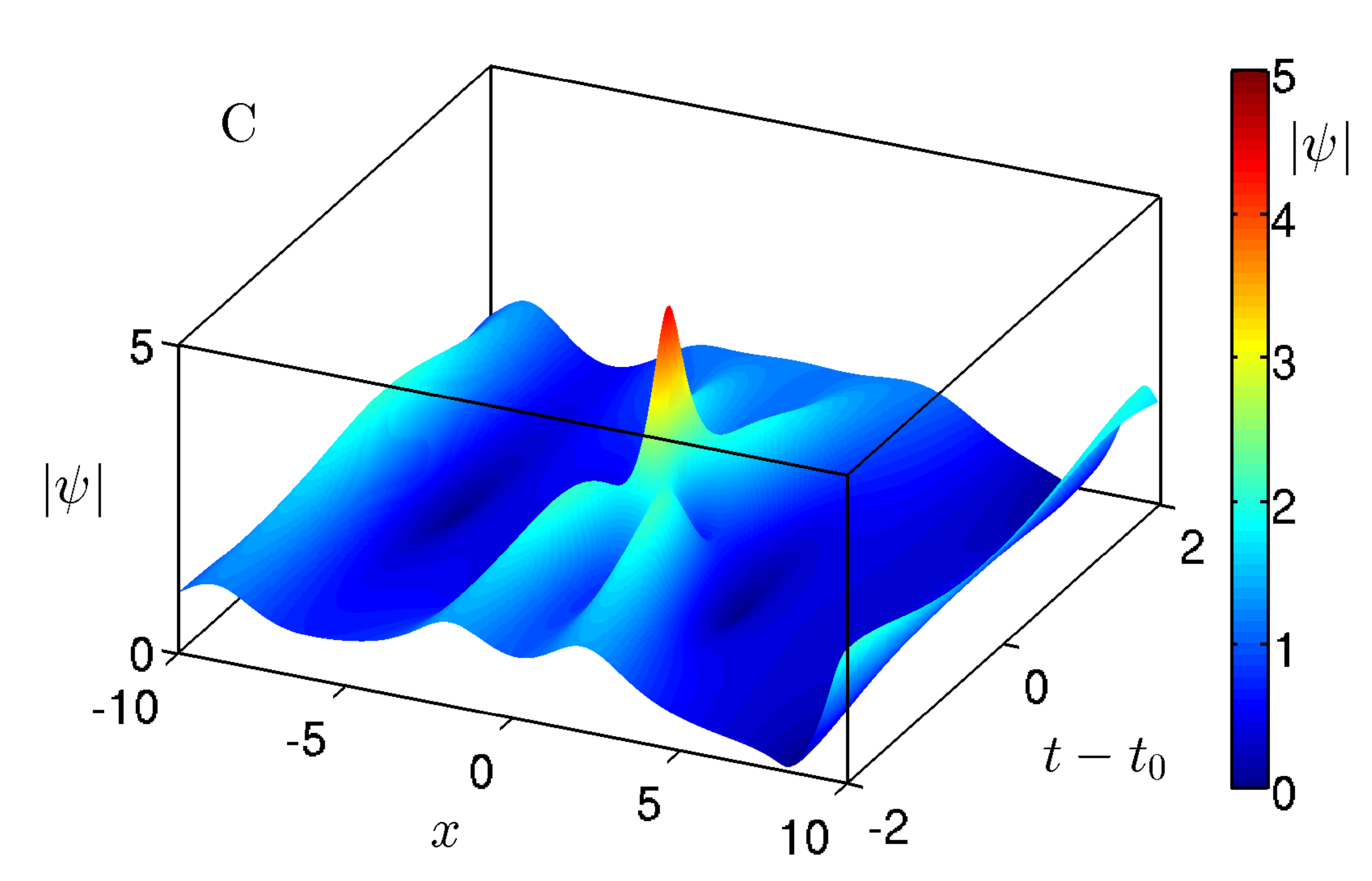}
	\includegraphics[width=8.8cm]{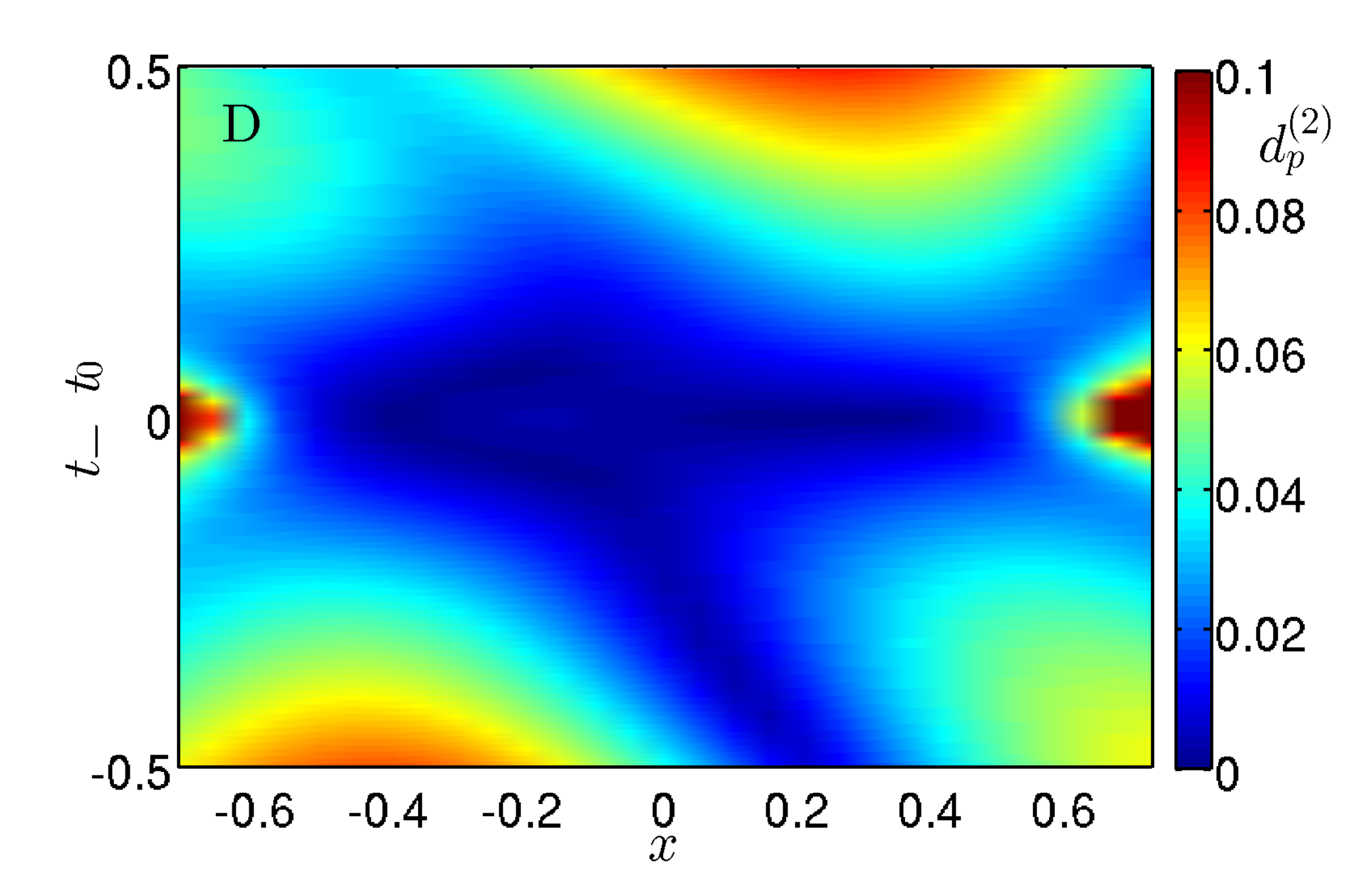}
	
	\caption{\small {\it (Color on-line)} One of the $10$ largest RWs (the coordinate of maximum amplitude is shifted to zero for better visualization) for the soliton gas case with time of occurrence $t_{0}\approx 39.2$, maximum amplitude $A\approx 4.4$ and deviation~(\ref{Grade}) from the RBS2 $\mathcal{D}_{p}^{(2)}\approx 0.02$: 
		\textbf{(A)} space profile of the RW $|\psi(x,t_{0})|$ at the time $t_{0}$ of its maximum elevation, 
		\textbf{(B)} time dependency of the maximum amplitude $\max_{x}|\psi|$, 
		\textbf{(C)} space-time representation of the amplitude $|\psi(x,t)|$ near the RW event, and 
		\textbf{(D)} relative deviation~(\ref{Grade-local}) between the wavefield and the fit with the RBS2 in the $(x,t)$-plane. 
		In the panel \textbf{(A)}, the thick black and thin dash-dot red lines indicate the space profile $|\psi(x,t_{0})|$ and the phase $\arg\psi(x,t_{0})$. 
		In the panels \textbf{(A,B)}, the dashed blue and green lines show the fits with the RBS1 and the RBS2, respectively. 
		In the panel \textbf{(D)}, the deviations $d_{p}^{(2)}\ge 0.1$ are demonstrated with constant deep red color. 
	}
	\label{fig:fig1}
\end{figure*}

An example of one of the $10$ largest RWs collected for the soliton gas case is shown in Fig.~\ref{fig:fig1}. 
The space profile $|\psi(x,t_{0})|$ and the phase $\arg\psi(x,t_{0})$ at the time of the maximum elevation $t_{0}\approx 39.2$ are demonstrated in Fig.~\ref{fig:fig1}(A), the temporal evolution of the maximum amplitude $\max_{x}|\psi|$ -- in Fig.~\ref{fig:fig1}(B), and the space-time representation of the amplitude $|\psi(x,t)|$ near the RW event -- in Fig.~\ref{fig:fig1}(C). 
As indicated in the figures, the space profile $|\psi(x,t_{0})|$ and the maximum amplitude $\max_{x}|\psi|$ are very well approximated by the amplitude-scaled RBS2, and the space-time representation strongly resembles that of the RBS2 as well. 
At the time of the maximum elevation, the RBS2 has four zeros; the RW presented in Fig.~\ref{fig:fig1} also has four local minimums that are very close to zero and where the phase $\arg\psi(x,t_{0})$ jumps approximately by $\pi$, see Fig.~\ref{fig:fig1}(A). 
Note that the phase is practically constant between the two local minimums closest to the maximum amplitude, as for the velocity-free RBS1 and RBS2. 
The described phase pattern is sometimes considered as a characteristic feature of RW formation, see~\cite{kedziora2013phase,xu2019phase}.

The deviation between a RW and its approximation with a RBS can be measured locally as
\begin{equation}\label{Grade-local}
	d_{p}^{(1,2)}(x,t) = \frac{|\psi - \psi_{p}^{(1,2)}|}{|\psi|}.
\end{equation}
Fig.~\ref{fig:fig1}(D) shows this deviation $d_{p}^{(2)}$ for the RBS2 in the $(x,t)$-plane: in space -- between the two local minimums closest to the maximum amplitude $x\in\Omega$, and in time -- in the interval $t-t_{0}\in[-0.5, 0.5]$, since outside the maximum amplitude noticeably deviates from the fit with the RBS2 in Fig.~\ref{fig:fig1}(C). 
The deviation $d_{p}^{(2)}$ remains well within 5\% for most of the area demonstrated in figure, so that the RBS2 turns out to be a very good approximation for the presented RW -- simultaneously in space and in time. 

As an integral measure reflecting the deviation between a RW and a RBS, one can consider the quantity
\begin{equation}\label{Grade}
	\mathcal{D}_{p}^{(1,2)} = \Bigg[\frac{\int_{x\in\Omega}\int_{t_{0}-\Delta T}^{t_{0}+\Delta T}|\psi - \psi_{p}^{(1,2)}|^{2}\,dx dt}{\int_{x\in\Omega}\int_{t_{0}-\Delta T}^{t_{0}+\Delta T}|\psi|^{2}\,dx dt}\Bigg]^{1/2}.
\end{equation}
Here we choose the region of integration over time $t\in[t_{0}-\Delta T, t_{0}+\Delta T]$ from the condition that at $t_{0}\pm\Delta T$ the RBS2 fit halves its maximum amplitude. 
Indeed, as demonstrated below, the collected RWs have maximum amplitudes roughly between $3.3$ and $5$, and their halving translates the waves below the RW threshold $|\psi| > 2.8$, see e.g.~\cite{agafontsev2015integrable}; also, for most of the RWs, the best fit is the RBS2. 
For the RW presented in Fig.~\ref{fig:fig1}, the interval of integration in time is $|t-t_{0}|\le 0.31$ and the deviations are $\mathcal{D}_{p}^{(1)}\approx 0.2$ for the RBS1 and $\mathcal{D}_{p}^{(2)}\approx 0.02$ for the RBS2. 

\begin{figure*}[t]\centering
	\includegraphics[width=8.8cm]{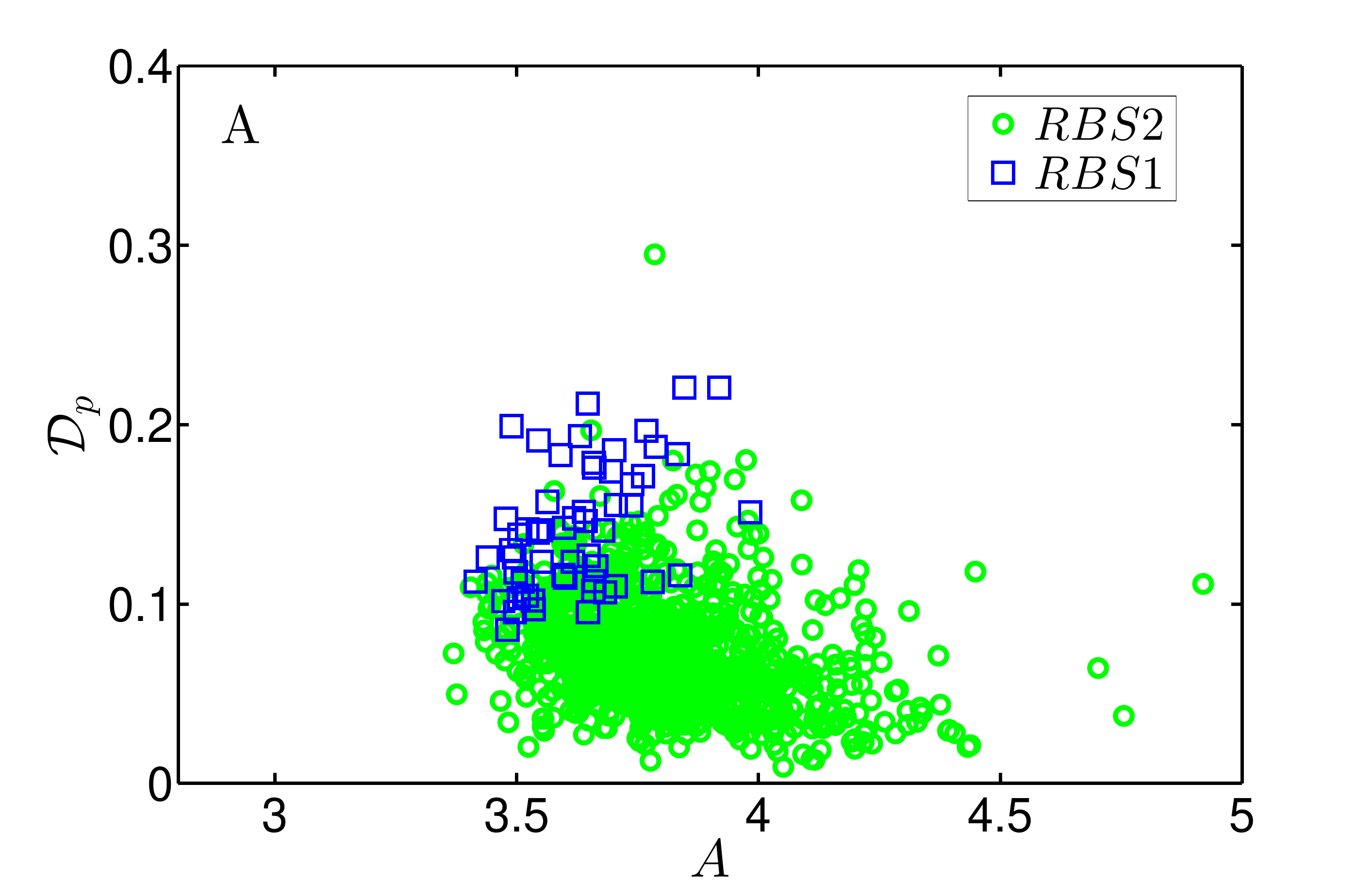}
	\includegraphics[width=8.8cm]{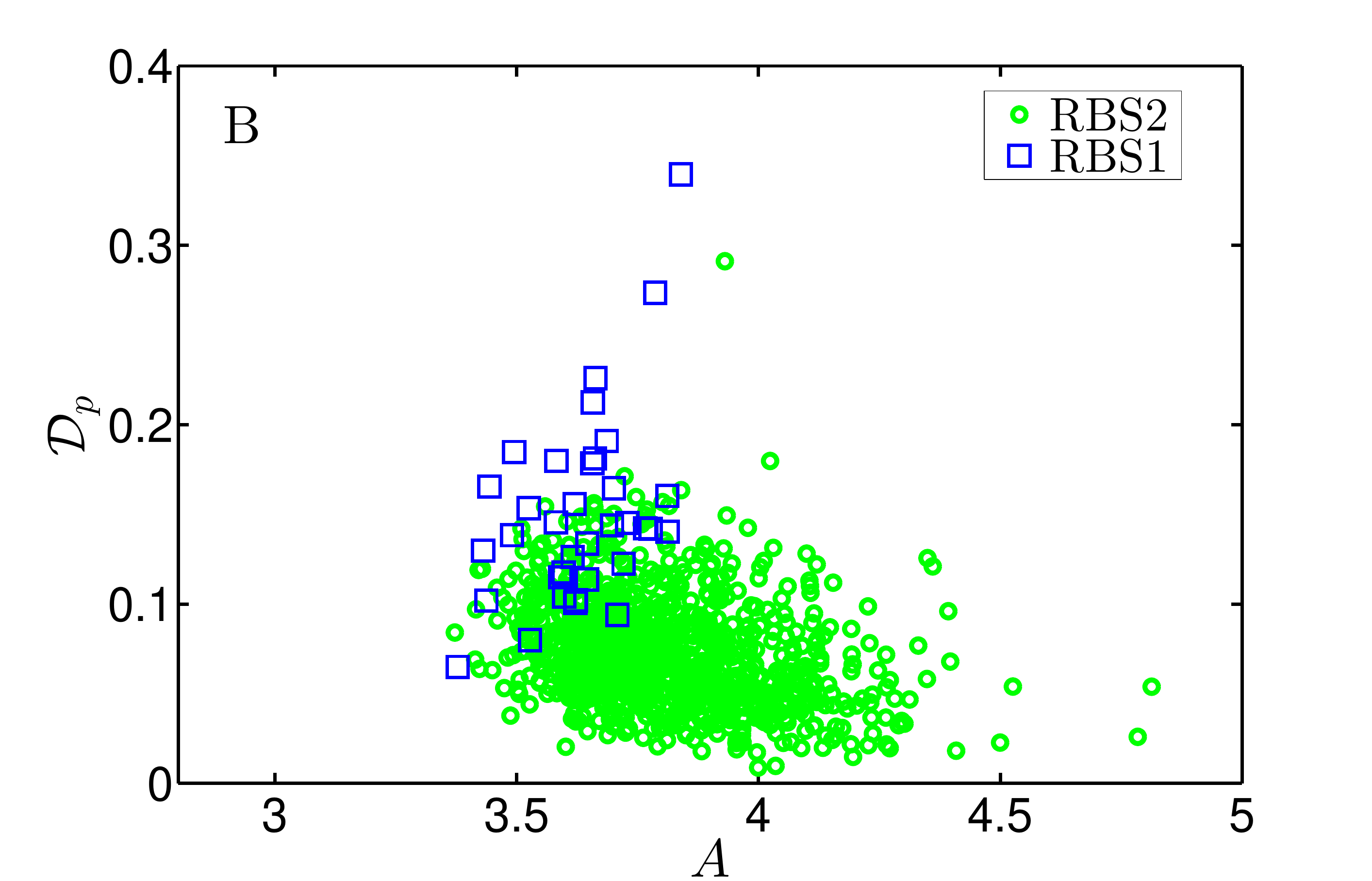}\\
	\includegraphics[width=8.8cm]{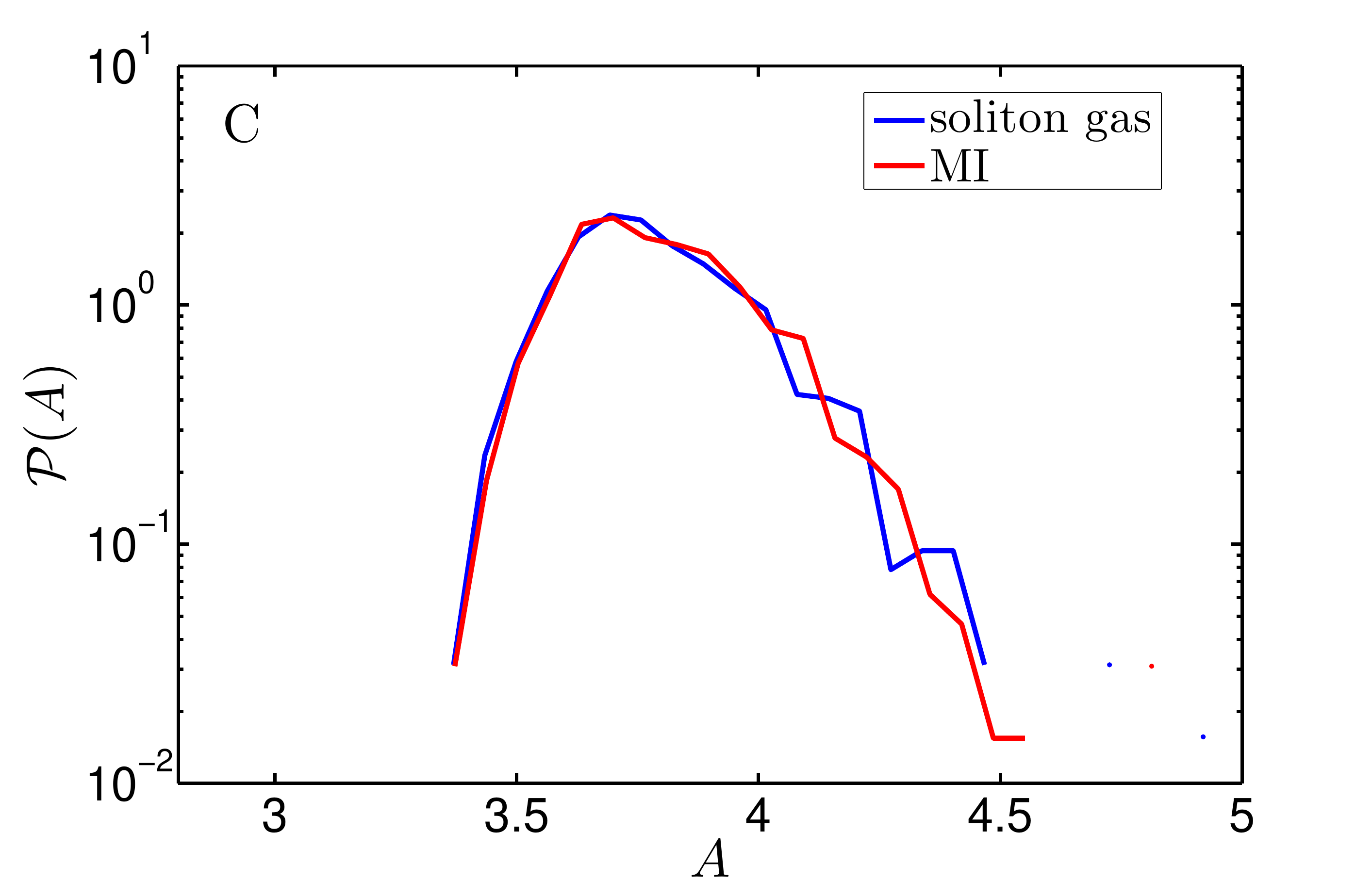}
	\includegraphics[width=8.8cm]{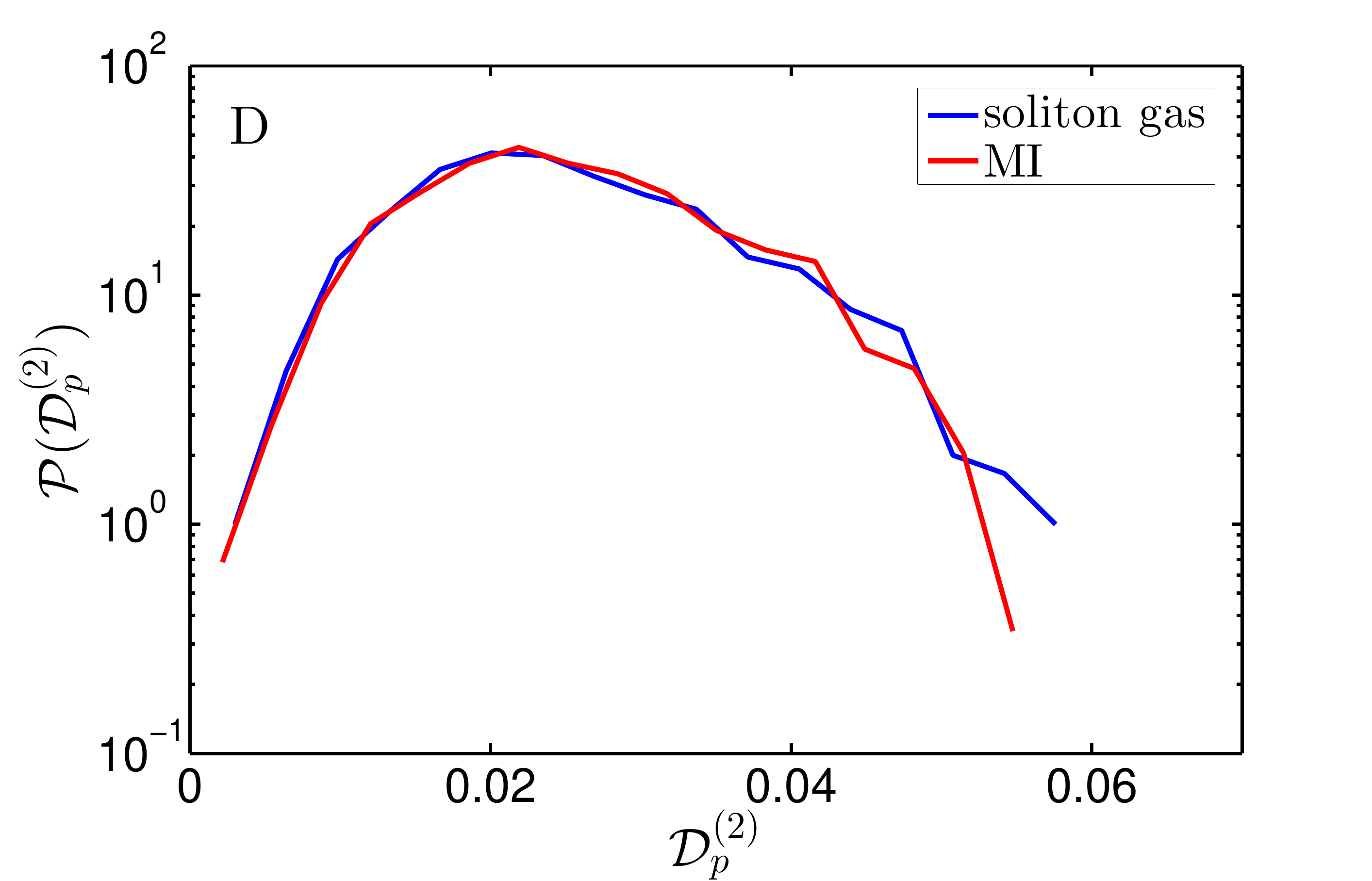}
	
	\caption{\small {\it (Color on-line)} \textbf{(A,B)} Deviation $\mathcal{D}_{p} = \min\{\mathcal{D}_{p}^{(1)}, \mathcal{D}_{p}^{(2)}\}$ between RWs and their best fits with either the RBS1 or the RBS2 versus the maximum amplitude $A$ of the RW: \textbf{(A)} for the soliton gas and \textbf{(B)} for the MI of the condensate close to its statistically stationary state. 
		The blue squares indicate that the best fit is achieved with the RBS1 and the green circles -- with the RBS2. 
		\textbf{(C,D)} The PDFs of \textbf{(C)} the maximum amplitude $A$ for all the RWs and \textbf{(D)} the deviation $\mathcal{D}_{p}^{(2)}$ for the RWs better approximated with the RBS2, for the soliton gas (blue) and the MI of the condensate close to its statistically stationary state (red).
	}
	\label{fig:fig2}
\end{figure*}

The quantity~(\ref{Grade}) can be used to assess how well a RW can be approximated by a RBS. 
Fig.~\ref{fig:fig2}(A) shows the minimum deviation $\mathcal{D}_{p}=\min\{\mathcal{D}_{p}^{(1)}, \mathcal{D}_{p}^{(2)}\}$ versus the maximum amplitude of the RW $A=\max|\psi|$, for all $1000$ RWs collected for the soliton gas case; the RWs better approximated with the RBS1 are indicated with blue squares and those with the RBS2 -- with green circles. 
For $57$ RWs the best fit turned out to be the RBS1 -- the Peregrine breather, while the other 943 RWs were better approximated by the RBS2. 
According to our observations, the value of deviation~(\ref{Grade}) below $0.05$ typically means that the RW is very well approximated with the corresponding RBS; for $0.05\lesssim \mathcal{D}_{p}^{(1,2)}\lesssim 0.1$ the approximation is satisfactory, and for $\mathcal{D}_{p}^{(1,2)}\gtrsim 0.1$ -- poor. 
Of $57$ RWs better approximated with the RBS1, only $4$ have deviations below $0.1$ and none -- below $0.05$; hence, the collected RWs can be approximated with the RBS1 satisfactory at best. 
For the RBS2 we have completely different picture: $768$ RWs show deviations from the RBS2 below $0.1$ and $220$ -- below $0.05$. 
As demonstrated in Fig.~\ref{fig:fig2}(A), larger RWs are typically better approximated with the RBS2. 
In particular, of $143$ RWs having maximum amplitude above $4$, $68$ have deviation from the RBS2 below $0.05$, and the mean deviation for the entire group of $143$ RWs is $\langle\mathcal{D}_{p}^{(2)}\rangle\approx 0.055$. 
Hence, we can conclude that the largest RWs are typically very well approximated by the RBS2. 

RWs collected close to the statistically stationary state of the noise-induced MI show the same general properties as those for the soliton gas case. 
Fig.~\ref{fig:fig2}(B) demonstrates very similar ``clouds'' of RWs approximated with either the RBS1, or the RBS2 on the diagram representing the minimum deviation $\mathcal{D}_{p}$ versus the maximum amplitude $A$. 
Of the $1000$ RWs in total, $36$ are better approximated with the RBS1 and $964$ -- with the RBS2. 
Of those better approximated with the RBS1, only $3$ have deviations below $0.1$ and none -- below $0.05$. 
Of $964$ RWs better approximated with the RBS2, $792$ have deviations below $0.1$ and $215$ -- below $0.05$. 
In total, $150$ RWs have amplitudes above $4$; out of them -- $64$ have deviation from the RBS2 below $0.05$, and the mean deviation among the group of $150$ RWs equals to $\langle\mathcal{D}_{p}^{(2)}\rangle\approx 0.059$. 

The RWs for the two systems turn out to be practically identically distributed by their maximum amplitude, as demonstrated in Fig.~\ref{fig:fig2}(C) with the corresponding probability density functions (PDFs). 
The PDFs of the deviation $\mathcal{D}_{p}^{(2)}$ for the RWs better approximated by the RBS2 (green circles in Fig.~\ref{fig:fig2}(A,B)) are also nearly identical, Fig.~\ref{fig:fig2}(D). 
Hence, we conclude that the largest RWs for the two systems show practically identical dynamical (resemblance with the RBS2) and statistical properties. 
Note that we have repeated simulations for the MI case with smaller and larger computational boxes and time windows for collecting the RWs. 
As a result, we have obtained the PDF of the maximum amplitude shifted to smaller or larger amplitudes, respectively. 
The nearly perfect correspondence of the two PDFs in Fig.~\ref{fig:fig2}(C) additionally justifies the usage of the simulation parameters discussed in the previous Section.


\section{Discussion}
\label{Sec:Discussion}

As we have mentioned in~\cite{gelash2018strongly}, some soliton collisions at the time of their maximum elevation have space profiles remarkably similar to those of the RBS1 and the RBS2. 
Moreover, we have presented an example of a phase-synchronized three-soliton collision, for which both the space profile and the temporal evolution of the maximum amplitude were very well approximated by the RBS2. 
The solitons in~\cite{gelash2018strongly} had nonzero velocities; here we modify the two- and three-soliton examples for the case of zero velocities and examine the local deviations $d_{p}^{(1,2)}(x,t)$~(\ref{Grade-local}) together with the integral deviations $\mathcal{D}_{p}^{(1,2)}$~(\ref{Grade}). 

\begin{figure*}[t]\centering
	\includegraphics[width=8.8cm]{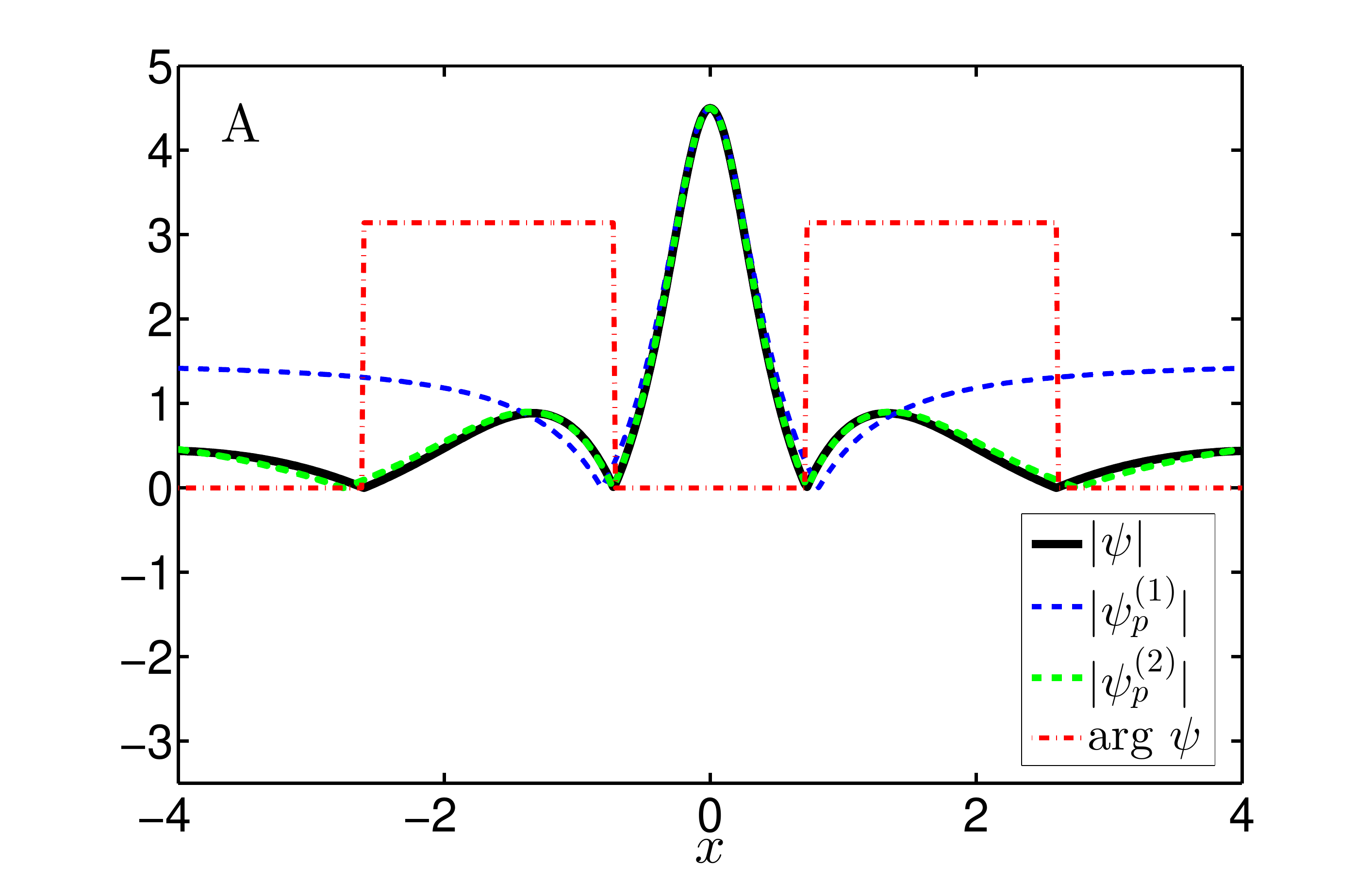}
	\includegraphics[width=8.8cm]{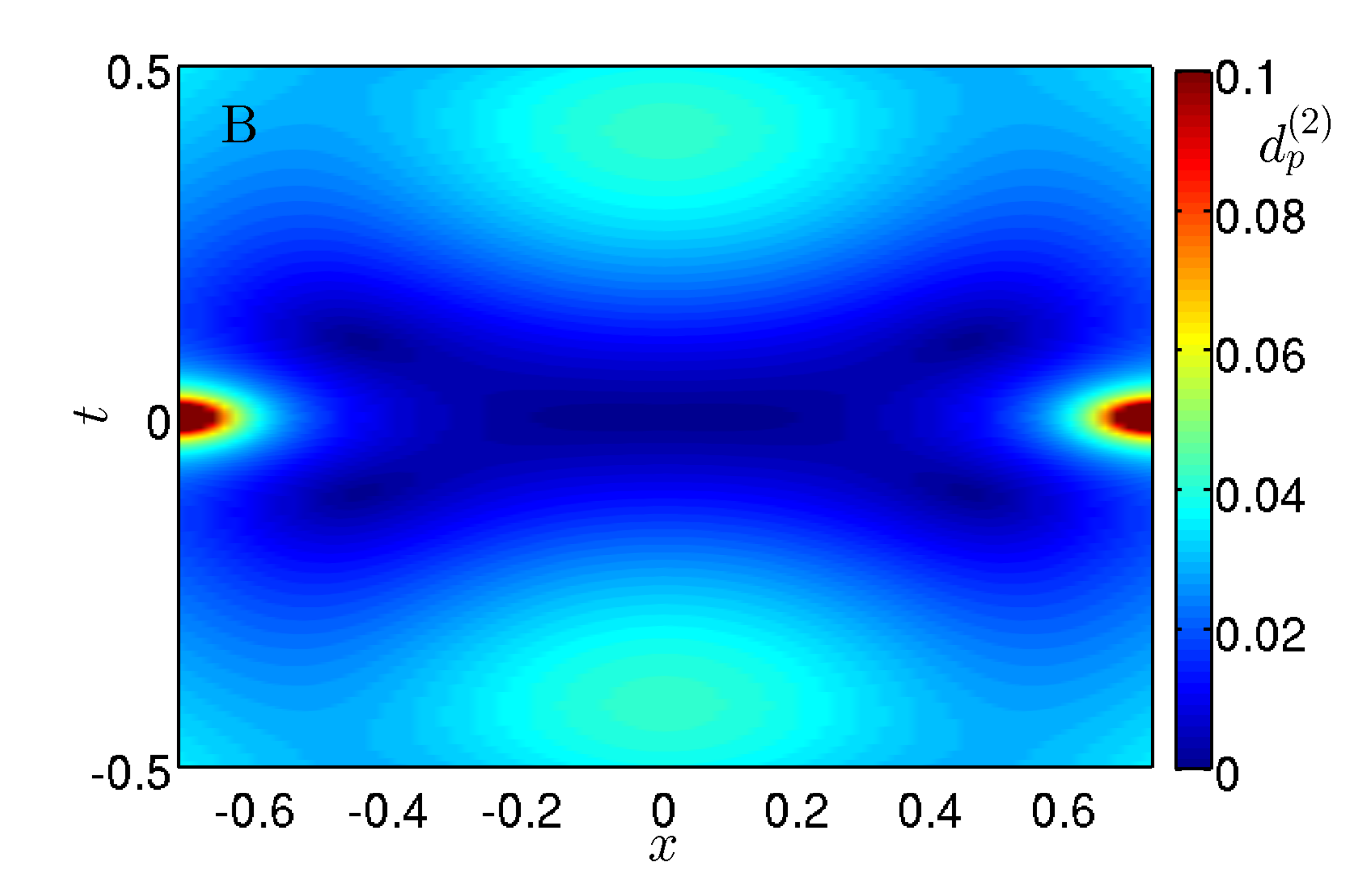}
	
	\caption{\small {\it (Color on-line)} Synchronized three-soliton interaction of solitons having amplitudes $a_{1} = 1$, $a_{2} = 1.5$ and $a_{3} = 2$, zero velocities $v_{j}=0$, zero space positions parameters $x_{0j}=0$ and, at the initial time $t=0$, zero phases $\Theta_{j}=0$: 
		\textbf{(A)} space profile $|\psi(x,t_{0})|$ and phase $\arg\psi(x,t_{0})$ at the time of the maximum elevation $t_0 = 0$, and 
		\textbf{(B)} relative deviation~(\ref{Grade-local}) between the wavefield and the fit with the RBS2 in the $(x,t)$-plane. 
		All notations are the same as in Fig.~\ref{fig:fig1}(A,D). 
		The deviation~(\ref{Grade}) from the RBS2 fit equals to $\mathcal{D}_{p}^{(2)}\approx 0.016$.
	}
	\label{fig:fig3}
\end{figure*}

Fig.~\ref{fig:fig3} shows an example of three-soliton interaction with solitons having amplitudes $a_{1} = 1$, $a_{2} = 1.5$ and $a_{3} = 2$, zero velocities $v_{j}=0$, zero space position parameters $x_{0j}=0$ and, at the initial time $t=0$, zero phases $\Theta_{j}=0$. 
The space profile $|\psi(x,t_{0})|$ at the time of the maximum elevation $t_0 = 0$ is remarkably similar to that of the RBS2, and the local deviation $d_{p}^{(2)}(x,t)$ remains well within $5\%$ for most of the area presented in the figure as well. 
The integral deviation~(\ref{Grade}) equals to $\mathcal{D}_{p}^{(2)}\approx 0.016$, that is even smaller than for the RW presented in Fig.~\ref{fig:fig1}. 

To analyze how often the phase-synchronized interactions of two and three solitons of various amplitudes may lead to such quasi-rational profiles, we have created $20$ two-soliton and $20$ three-soliton interactions with solitons of random amplitudes, zero velocities $v_{j}=0$, zero space positions parameters $x_{0j}=0$ and phases $\Theta_{j}=0$. 
For the two-soliton interactions, the minimum deviations from the RBS1 and the RBS2 turned out to be $\mathcal{D}_{p}^{(1)}\approx 0.077$ and $\mathcal{D}_{p}^{(2)}\approx 0.061$, and the average ones -- $\langle\mathcal{D}_{p}^{(1)}\rangle\approx 0.14$ and $\langle\mathcal{D}_{p}^{(2)}\rangle\approx 0.075$, respectively. 
For the three-soliton case, the minimum deviations were $\mathcal{D}_{p}^{(1)}\approx 0.18$ and $\mathcal{D}_{p}^{(2)}\approx 0.003$, and the average ones -- $\langle\mathcal{D}_{p}^{(1)}\rangle\approx 0.23$ and $\langle\mathcal{D}_{p}^{(2)}\rangle\approx 0.022$; the maximum deviation from the RBS2 equaled to $\mathcal{D}_{p}^{(2)}\approx 0.03$, that is still very good for comparison with the RBS2. 
Hence, we conclude that quasi-rational profiles very similar to that of the RBS2 appear already for three-soliton interactions, provided that the solitons are properly synchronized (that is, have coinciding positions and phases). 

We think that the presented elementary three-soliton model might provide an explanation of RW formation inside multi-soliton solutions. 
The most direct way for future studies might be a demonstration of a RW for synchronized many-soliton solution. 
Here, however, we face a new question, that is, whether formation of a RW is a collective phenomenon that requires synchronization of all the solitons, or a ``local'' event that can be achieved by synchronizing of a few. 
Note that even the latter case represents a challenging problem. 
Indeed, the solitons generating a RW acquire space and phase shifts due to presence of the remaining solitons, that should influence their optimal synchronization condition. 
For remote solitons, the shifts can be computed analytically using the well-known asymptotic formulas, see e.g.~\cite{novikov1984theory}, which however do not work for our case of a dense soliton gas where all solitons effectively interact with each other. 
This leaves us two options: (i) the local numerical synchronization of a small group with ``trial and error'' method and (ii) the calculation of the generalized space-phase shifts expressions for the closely located solitons.

Also note that our study is limited with respect to statistical analysis of RWs, as we have focused on the largest RWs, while the ``common'' RWs may have different dynamical and statistical properties. 
Nevertheless, we believe that, since the largest RWs for the two systems show identical properties, the ``common'' RWs have the same properties too. 
Identification of all the RWs according to the standard criterion $A\ge 2.8$ is a nontrivial problem by itself, and we plan to return to it in the near future.


\section{Conclusions}
\label{Sec:Conclusions}

In this brief report we have presented our observations of RWs within the 1D-NLSE model for (i) the modulationally unstable plane wave at its long-time statistically stationary state and (ii) the bound-state multi-soliton solutions representing the solitonic model of this state. 
Focusing our analysis on the largest RWs, we have found their practically identical dynamical and statistical properties for both systems. 
In particular, most of the RWs turn out to be very well approximated -- simultaneously in space and in time -- by the amplitude-scaled rational breather solution of the second order (RBS2), and the two sets of the collected RWs are identically distributed by their maximum amplitude and deviation from the RBS2. 
Additionally, we have demonstrated the appearance of quasi-rational profiles very similar to that of the RBS2 already for synchronized three-soliton interactions. 

The main messages of the present paper can be summarized as follows. 
First, a quasi-rational profile very similar to a RBS does not necessarily mean emergence of the corresponding rational breather, as it can be a manifestation of a multi-soliton interaction. 
Second, the identical dynamical and statistical properties of RWs collected for the two examined systems strongly suggest that the main mechanism of RW formation should be the same, i.e., that RWs emerging in the asymptotic stationary state of the MI (and, possibly, in other strongly nonlinear wavefields) are formed as interaction of solitons. 
However, more study is necessary to clarify how exactly interaction of solitons within a large wavefield may lead to formation of a RW, and we plan to continue this research in future publications.


\begin{center}
\textbf{Acknowledgements}
\end{center}

Simulations were performed at the Novosibirsk Supercomputer Center (NSU). 
The work of D.S.A was supported was supported by the state assignment of IO RAS, Grant No. 0149-2019-0002. 
The work of A.A.G was supported by RFBR Grant No. 19-31-60028.


%

\end{document}